\documentclass[cits]{PoS}

\def\HI{H{\,\small I}}

\title{Mid-frequency aperture arrays: the future of radio astronomy}

\ShortTitle{Mid-frequency aperture arrays}

\author{\speaker{Ilse M. van Bemmel}\\
        Netherlands Institute for Radio Astronomy (ASTRON), PO Box 2, 7900 AA Dwingeloo, The~Netherlands\\
        E-mail: \email{bemmel@astron.nl}
        }

\author{Arnold van Ardenne\\
        Netherlands Institute for Radio Astronomy (ASTRON)
}

\author{Jan Geralt bij de Vaate\\
        Netherlands Institute for Radio Astronomy (ASTRON)
}

\author{Andrew J. Faulkner\\
	University of Cambridge, Cavendish Laboratory, JJ Thompson Avenue, Cambridge, CB3 0HE, United Kingdom
}

\author{Raffaella Morganti\\
        Netherlands Institute for Radio Astronomy (ASTRON)
}

\abstract{Aperture array (AA) technology is at the forefront of new developments and discoveries in radio astronomy. Currently LOFAR is successfully demonstrating the capabilities of dense and sparse AA's at low frequencies. For the mid-frequencies, from 450 to 1450\,MHz, AA's still have to prove their scientific value with respect to the existing dish technology. Their large field-of-view and high flexibility puts them in an excellent position to do so.
The Aperture Array Verification Program is dedicated to demonstrate the feasibility of AA's for science in general and SKA in particular. For the mid-frequency range this has lead to the development of EMBRACE, which has already demonstrated the enormous flexibility of AA systems by observing \HI\ and a pulsar simultaneously. It also serves as a testbed to demonstrate the technological reliability and stability of AA's. The next step will put AA technology at a level where it can be used for cutting-edge science. 
In this paper we discuss the developments to move AA technology from an engineering activity to a fully science capable instrument. We present current results from EMBRACE, ongoing tests of the system, and plans for EMMA, the next step in mid-frequency AA technology.}

\FullConference{Resolving the Sky - Radio Interferometry: Past, Present and Future -RTS2012\\
		April 17-20, 2012\\
		Manchester, UK}

\begin{document}

\begin{figure}
\centering
\includegraphics[width=12cm]{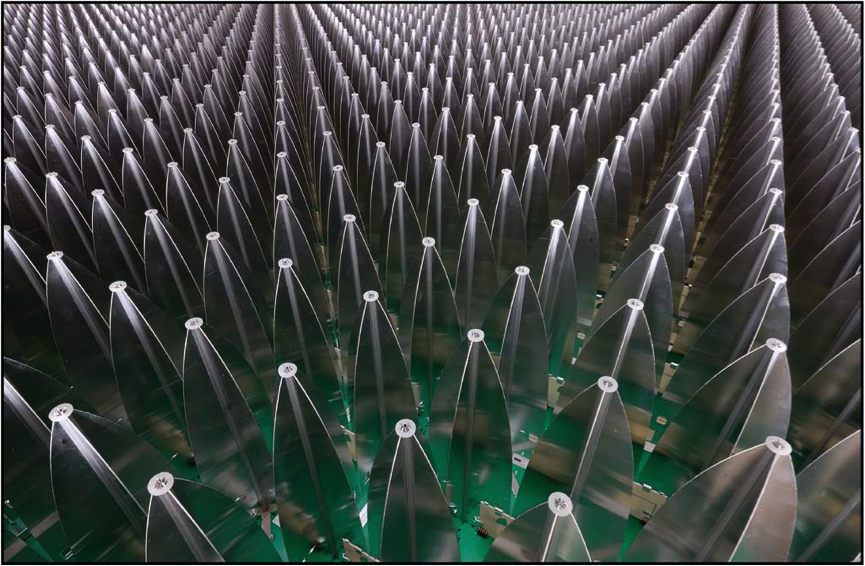}
\caption{Actual image of a part of one of the EMBRACE stations.}
\label{embrace}
\end{figure}

\section{Introduction: dense aperture arrays concept}
Aperture arrays (AA's, see Fig.~\ref{embrace}) have been under development at ASTRON for decades, and the technology is now reaching a readiness level at which it can be applied in the construction of future scientific instruments, most notably the Square Kilometre Array (SKA). In this paper we discuss the so-called 'dense' aperture arrays, where the individual receiver elements are closely spaced, and the system is optimized to observe at a frequency range from roughly 450 to 1450\,MHz. This technology is part of the Advanced Instrumentation Programme for the SKA.

Dense AA's have many advantages over dishes. The individual elements are intrinsically broad band receivers, e.g. for Vivaldi elements a frequency range of 9:1 is easily achievable, therefore these are a strong candidate for the individual antenna elements, see also \cite{andy_talk}. Due to computational limitations at these higher frequencies, the operational range required is set to a more feasible 3:1. Since AA's have no moving parts, they can be repointed in a matter of seconds. This makes the technology extremely suitable for rapid follow-up of transient events. 

A unique feature of AA's is the possibility to observe in multiple directions simultaneously (see also Fig.~\ref{emma_multibeam}). With the field-of-view in each direction being an order of magnitude larger than that for dishes with single pixel feeds, AA's form an ideal technology to build a cheap and highly efficient survey instrument. The size of the field-of-view is in principle only limited by the available processing power, creating the potential of unprecedented survey speeds. In order to exploit this, cost reduction of beam forming and post-processing is the main challenge. 

AA's employ multi-level beam forming. First, the wide beams of individual receiver elements are combined into a smaller beam, which is the equivalent of a primary beam in a dish. This defines the field-of-view of the instrument, and we will refer to it as such. Subsequently, all available fields are digitally combined to achieve the highest possible spatial resolution. We will refer to this as a beam. 

\section{EMBRACE}
EMBRACE was built as part of the SKA Design Study (SKADS) \cite{skads_overview09, skads_whitepaper}. There is a station at the location of WSRT, and one at Nan\c{c}ay in France \cite{embrace_tap}. Over the PrepSKA years (2008- April 2012), the development of EMBRACE was continued under the Aperture Array Verification Programme \cite{aavp09}. Towards the end of this year (2012), the EMBRACE station at WSRT will be fully constructed. The two analogue fields-of-view are now fully operational, using entirely separated signal paths. Although the tiles are fitted with dual polarization receptors, only a single polarization is connected and processed.

With their limited sensitivity, each EMBRACE station is too small for ground-breaking astronomical observations, though some basic tests have been performed using bright sources such as satellites, the Sun and the brightest pulsar. Additional verification observations are being designed and planned for this year.

\begin{figure}
\centering
\includegraphics[width=7cm]{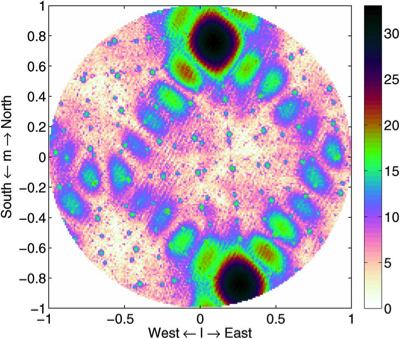}
\includegraphics[width=7.3cm]{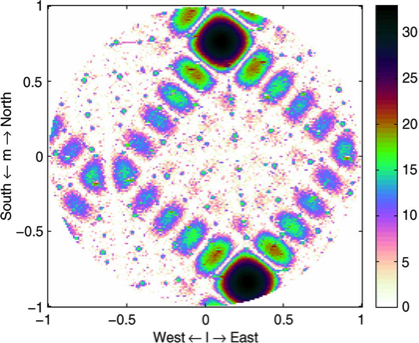}
\caption{{\it Left}: Observation of the Afristar satellite with 3x3 tiles. Due to the low elevation, a grating lobe is visible on the northern horizon. {\it Right:} simulation of the expected beam pattern for the observation. Note the close similarity between model and observation of the small peaks between the two major grating lobes. Colour scales are in decibel.}
\label{afristar}
\end{figure}

\begin{figure}
\includegraphics[width=5cm]{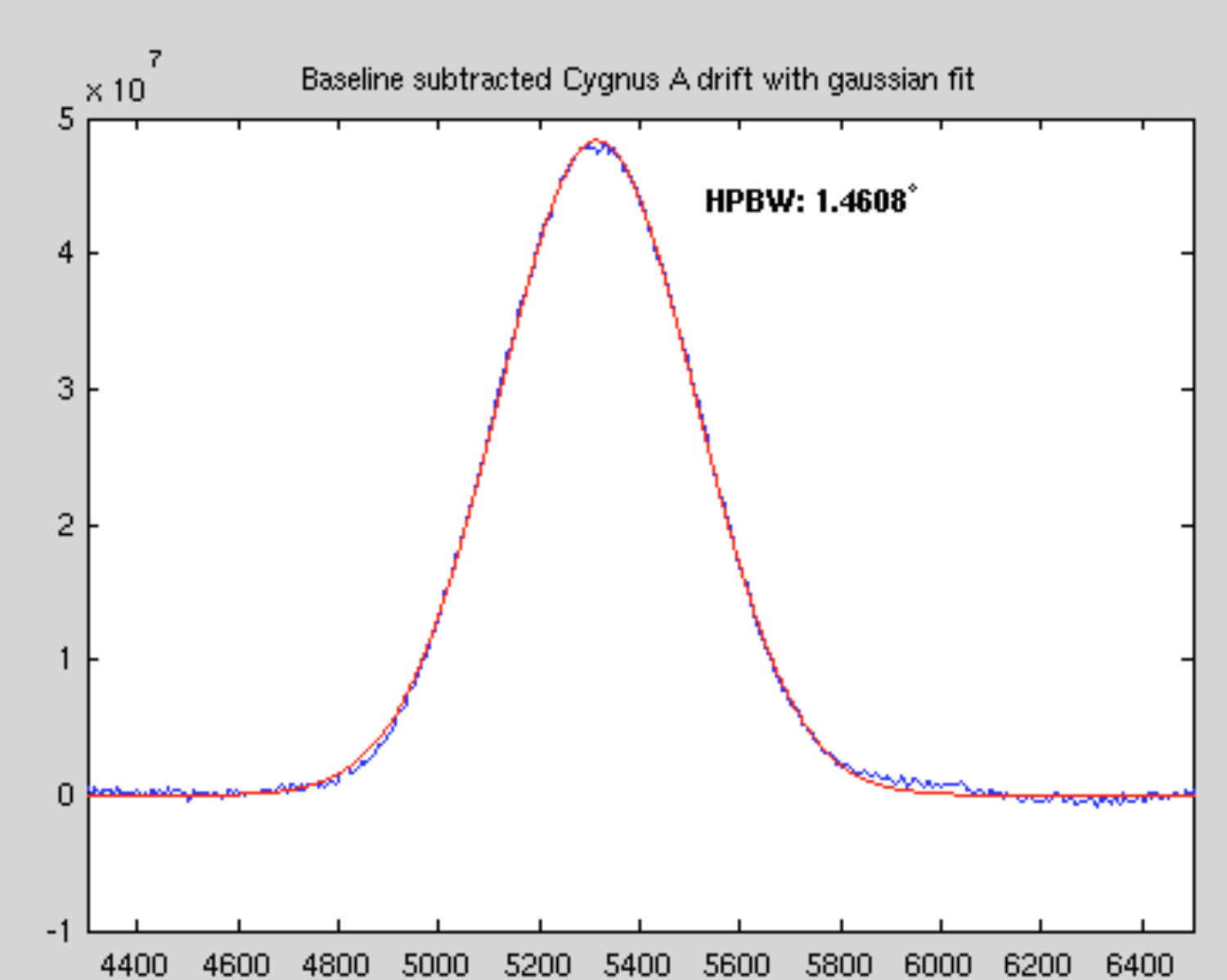}
\includegraphics[width=10cm]{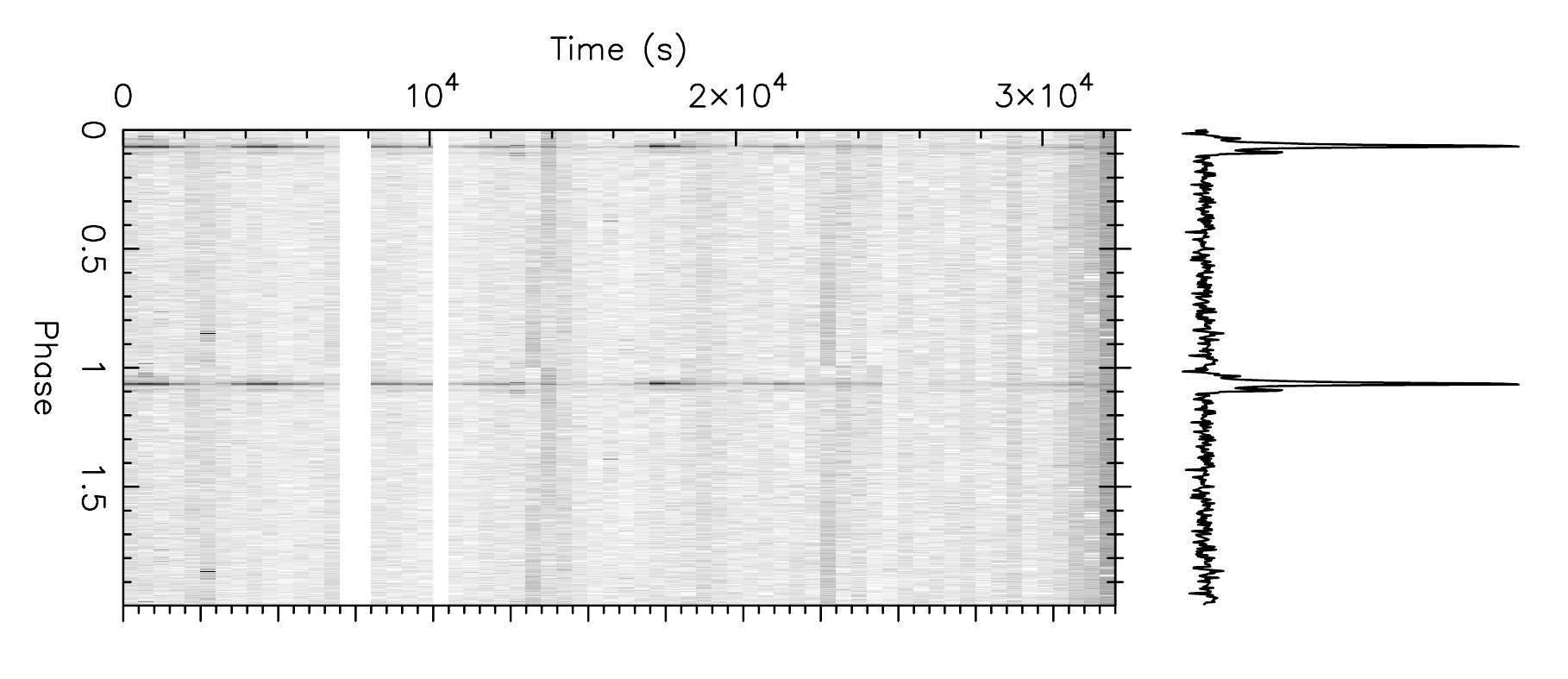}
\caption{{\it Left}: Driftscan of Cygnus~A. Blue is the observation, red is the model. Observation done with the Nan\c{c}ay EMBRACE station. {\it Right:} Continuous observation over 9 hours of pulsar B0329+54. White regions are due to RFI mitigation. Observation done with the WSRT EMBRACE station.}
\label{drift_pulsar}
\end{figure}

The main results demonstrated with EMBRACE include:
\begin{enumerate}
\item {\it Dual field observations of a pulsar and all-sky \HI\ map}. This experiment demonstrates the feasibility of doing two entirely different observations {\em simultaneously}. While one field-of-view was tracking a pulsar, the other was mosaicing the sky.
\item {\it Beam profile on Afristar satellite}. For a 3x3 tile configuration an observation was done on the geostationary Afristar satellite. The results were compared with a model of the expected beam, and agreed to within the 1\% level. This demonstrates the high accuracy of beam forming which is achieved (see Fig.~\ref{afristar}), for more details see \cite{wijnholds_embrace09}.
\item {\it Drift scan experiments on bright continuum sources}. Cygnus~A is the brightest non-varying continuum source in the Northern sky, and is used by the Nan\c{c}ay station to make drift scans. After subtraction of the Milky Way background, the beam corresponds closely to the model prediction. However, Cygnus~A is not strong enough to reveal the first sidelobe (see Fig.~\ref{drift_pulsar}).
\item {\it Repeated observations of the brightest pulsar}. Multiple observations of pulsar B0329+54 have been obtained. The pulsar can be tracked for over 9 hours after a \em{single} calibration (see Fig.~\ref{drift_pulsar}). These repeated experiments demonstrate the system stability.
\end{enumerate}

The results so far are only the beginning of many more extensive tests. With two stations, either station can be used to corroborate the results of the other. The pen-ultimate test will be to obtain fringes between the WSRT and Nan\c{c}ay stations. It is also being explored to combine the WSRT station with one or more WSRT dishes for a combined imaging experiment. Other tests that are ongoing, to be done, or repeated with the completed system include:
\begin{enumerate}
\item {\it Field switching experiments}. With the two fields-of-view now using fully separated paths, we can perform switching experiments, in which the fields-of-view are switched between two bright sources at a regular time interval. This will reveal how stable and identical the gains within the fields-of-view are (a.k.a. the stability of the primary beam).
\item {\it Detect second brightest pulsar}. As we have a measurement of A/T, we can predict how much time it will take to detect the second brightest pulsar in the sky visible to EMBRACE. The calculation predicts $\sim1$ hour. This will be verified with an actual observation.
\item {\it Track and observe bright continuum sources}. Last December a major bug was fixed which prevented tracking of the pulsar for more than 1 hour. Now that we are confident the tracking works for pulsars, the next step will be to track continuum sources and verify the tracking stability.
\item {\it Deep field (cold sky) observations}. For spectral observations a smooth bandpass is essential. This can be verified by observing an area of cold sky. If there are instrumental resonances, they will show up as a peak in the auto correlation function at their specific frequency.
\item {\it \HI\ observations of (extra-)galactic sources}. Observation of \HI\ in the Galaxy, and in nearby galaxies, will be a thorough system test at all levels. Potential sources for this include M33, M42, but also the high-velocity clouds near our Galaxy.
\end{enumerate}

From these tests we get a good handle on the overall system stability. The shape of the primary beam should be characterized to a high precision. Sensitivity, bandpass stability, and aperture efficiency will be measured and matched to the design specifications. 

\section{The next step: EMMA}
Though EMBRACE is a necessary and useful testbed for AA instruments, it has its limitations, mainly due to its size and lack of polarization. To bridge the gap between EMBRACE and the requirements for AA's in SKA, a science capable demonstrator is needed. This will be EMMA, whose name is more of an actual name than an acronym. Besides being a verification system as seen from SKA, it is also a science capable pathfinder mission. 

\begin{figure}
\centering
\includegraphics[width=12cm]{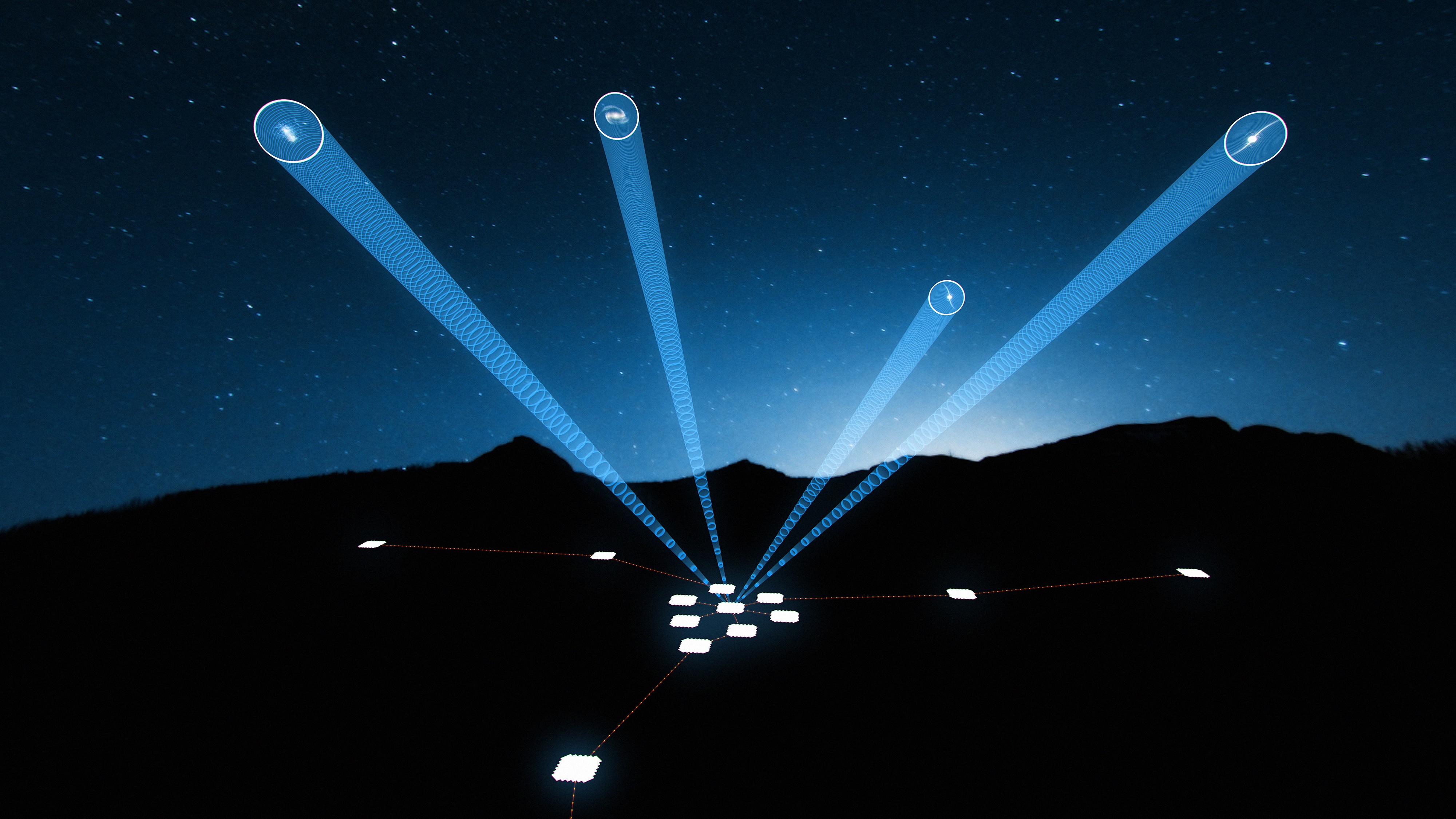}
\caption{Artist's impression of a multi-beam observation with EMMA. A field-of-view is rendered for each single polarization, yielding a total of four independent fields-of-view (image credit: Swinburne Astronomy Productions).}
\label{emma_multibeam}
\end{figure}

The scientific goal of EMMA is to detect the Baryonic Acoustic Oscillations (BAO) through \HI\ intensity mapping \cite{bao_intensitymapping}. In order to do so, the technology will have to meet the current SKA requirements. As a consequence, EMMA will be a powerful instrument in its own right, capable of brand new studies of polarization, detection of large scale \HI, \HI\ absorption up to $z\sim2.5$, pulsars, and transients studies.

The technical goal is to demonstrate the potential for interferometric imaging. EMMA will reveal the unique potential and feasibility of AA systems for astronomical purposes, such as multi-directional observations, large field-of-view and rapid response time at these frequencies. As with LOFAR and other AA systems, EMMA will provide the similar fabulous flexibility far beyond any dish capability.

As part of the SKA development, EMMA will explore options to operate fully on renewable energy, and use biodegradable materials where possible. Funding opportunities along these lines are being explored. Further development of the exploitation of green energy is of high importance for the operation of SKA.

\section{EMMA design and configuration}
In the current design EMMA will be a 14 station interferometer with a maximum baseline of about 1\,km. The details of the design are listed in Table~\ref{emma_specs}. The station size is limited by the field-of-view requirement for the BAO science case. The most notable aspect of EMMA will be the multiple fields-of-view (see Fig.~\ref{emma_multibeam}). This unique capability will allow for entirely new observational methods. EMMA will be observing in a continuous frequency range from $\sim$450$-$1450\,MHz, where none of the existing observatories can observe more than a narrow band at a fixed frequency.

\begin{table}
\centering
\begin{tabular}{lc}
{\bf Design parameter} 	& {\bf Value} \\
\hline
Maximum longest baseline	& $<$1\,km \\
Frequency range			& 450$-$1450\,MHz \\
Total collecting area			& 2000\,m$^2$ \\
Number of stations		& 14 \\
Spectral resolution			& 4\,kHz \\
Field-of-view per station at 1.4\,GHz & 78\,deg$^2$\\
Instantaneous bandwidth		& 500\,MHz \\
System temperature			& $<$50\,K \\
Polarization per field-of-view	& Full Stokes \\
Independent fields-of-view	& 2 \\
Time resolution				& 50\,$\mu$s \\
Electronic scan angle per receiving element & $\pm45^\circ$ \\
Digital beams within analogue beam & 64\\
\hline
\end{tabular}
\caption{Current design specifications for EMMA}
\label{emma_specs}
\end{table}

\begin{figure}
\centering
\fbox{\includegraphics[width=7cm]{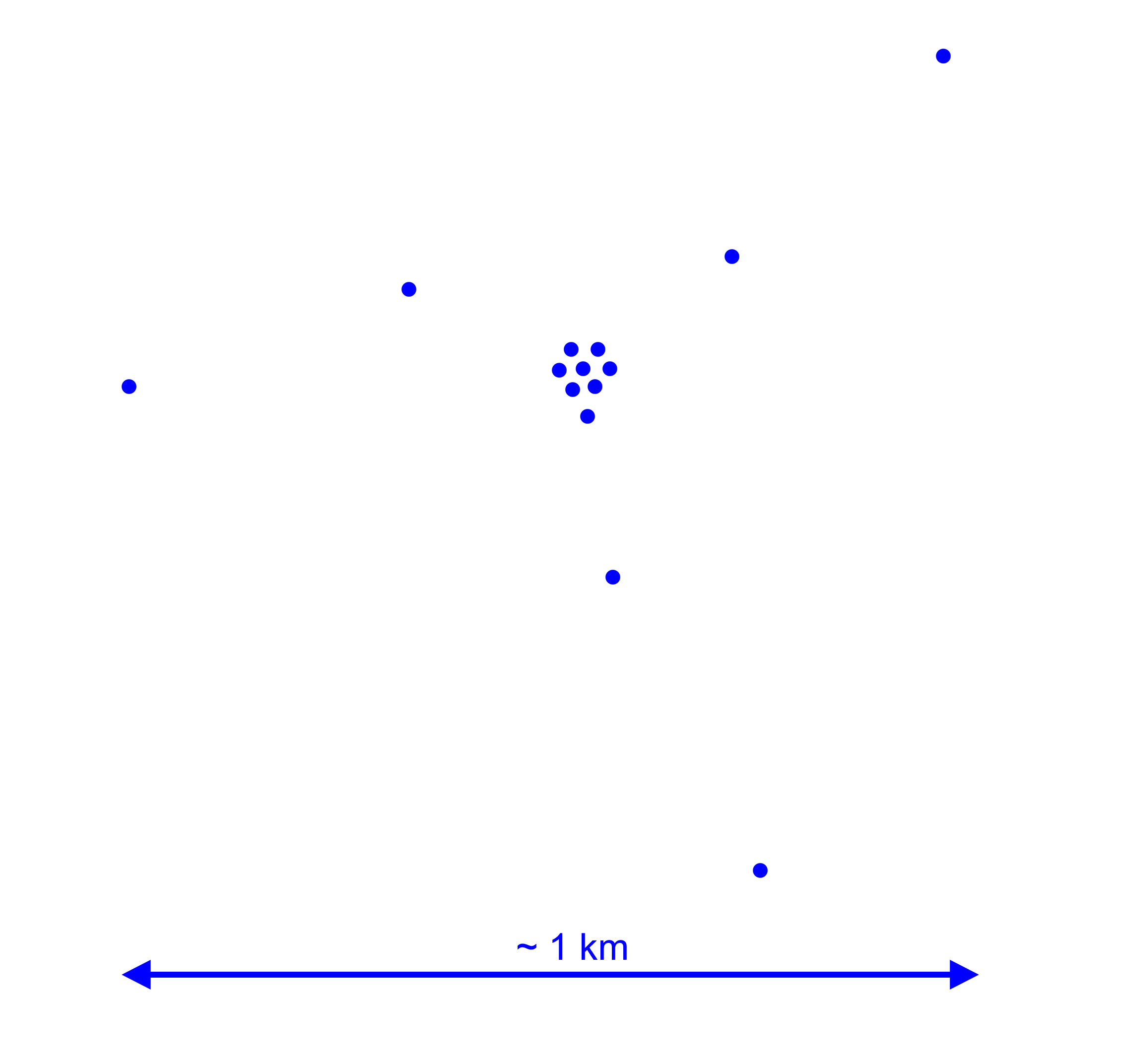}}
\caption{Current plan for the EMMA station layout. The size of each circular station is roughly to scale. In practice, each station may look more square due to the 1\,m$^2$ size of the individual tiles which harbour the receivers.}
\label{emma_layout}
\end{figure}

The configuration of EMMA will be a compact core with $>$50\% of the collecting area inside a region of 100\,m, comparable to the LOFAR superterp. In addition, there will be several stations positioned at distances of a few hundred metres from the core, either individually, or along 'arms' in analogy to existing interferometers (see Fig.~\ref{emma_layout}). Though the core in Fig.~\ref{emma_layout} is represented by individual stations, in practice it could be a single large station, which is electronically divided into smaller units.

\section{Planning}
The technical development for EMMA will require improvements at tile level. EMBRACE is already fitted with dual polarization antennas, so the main development for EMMA will be to lower the costs (making the tiles mass produceable), and implement signal paths which can cope with full polarization and larger bandwidths. This can be completed on a relatively short timescale, with the first EMMA tiles being tested during 2013, and first (sub-)station tests in 2014. To keep in line with the SKA timeline, the goal is to have at least 4 EMMA stations of 100\,m$^2$ each operational before decisions are made on the Advanced Instrumentation Programme for SKA phase 2, currently scheduled for 2016. The full EMMA will be operational in parallel to SKA phase 1.

\section{Summary}
The testing and verification of EMBRACE has delivered valuable insights, as well as ground-breaking results. We have successfully observed simultaneously in two independent directions, and achieved the required high accuracy in beam forming and calibration. With both stations operational, additional tests will continue throughout this year and into the next. In parallel, the development of improved tiles will take place for the next step: EMMA. 

EMMA will be a science capable AA instrument, focussing on detecting the BAO signal and demonstrating imaging capabilities. Though it will be small, it can match survey speed with most of the upcoming and existing radio observatories in the mid-frequencies. Its multiple, large fields-of-view help to overcome the limited point-source sensitivity. Most notably, EMMA will provide access to a unique and continuous frequency range, which is only marginally covered by existing observatories. With the current specifications, EMMA will be a powerful and very flexible system, capable of a large range of scientific observations. It will make important progress in our understanding of the use of AA's in a scientific instrument, a necessary step in the evolution of AA technology towards the SKA.

\subsection*{Acknowledgements}
The authors would like to thank all the people involved in the construction, testing and operation of EMBRACE. Most notably, we thank Steve Torchinsky for sharing with us the results of the Nan\c{c}ay EMBRACE station. Thanks is also due to the people in the astronomy group at ASTRON for extensive discussion on the capabilities of EMMA. We thank Russ Scott at Swinburne Astronomy Productions for a very pleasant collaboration in making the animation and images.

\bibliographystyle{amsplain}
\bibliography{AA}

\end{document}